# GMOL: An Interactive Tool for 3D Genome Structure Visualization

**Jackson Nowotny[1], Avery Wells[1], Lingfei Xu[1], Renzhi Cao[1], Tuan Trieu[1], Chenfeng He[1], Jianlin Cheng[1, 2, 3]***

[1]Computer Science Department, University of Missouri, Columbia, MO 65211, USA

[2]Informatics Institute, University of Missouri, Columbia, MO 65211, USA

[3]C.S. Bond Life Science Center, University of Missouri, Columbia, MO 65211, USA

*Corresponding author

Email addresses:

    AW: aowhd4@mail.missouri.edu

    JN: jn9qc@mail.missouri.edu

    LX: lxq35@mail.missouri.edu

    RC: rcrg4@mail.missouri.edu

    TT: tuantrieu@mail.missouri.edu

    CH: chbn6@mail.missouri.edu

    JC: chengji@missouri.edu

# Abstract

**Background**

It has been shown that genome spatial structures largely affect both genome activity and DNA function. Knowing this, many researchers are currently attempting to accurately model genome structures. Despite these increased efforts there still exists a shortage of tools dedicated to visualizing the genome. Creating a tool that can accurately visualize the genome can aid researchers by highlighting structural relationships that may not be obvious when examining the sequence information alone. Here we present a desktop application, known as GMOL, designed to effectively visualize genome tertiary structures at multiple scales so that researchers may better analyze their genomic data.

**Results**

GMOL was developed based upon our multi-scale approach that allows a user to zoom in and out between six separate levels within the genome. These six scales are full genome, chromosome, loci, fiber, nucleosome, and nucleotide. In order to store the data of the different scales, a new file format, known as GSS, was created. With GMOL, a user can choose any unit at any scale and scale it up or down to visualize its structure and retrieve corresponding genome sequences from either Ensembl or a local database. Users can also interactively manipulate and measure the whole genome structure and extract static images and machine-readable data files in PDB format from the multi-scale structure.

**Conclusion**

By using GMOL researchers will be able to better understand and analyze genome structure models and the impact their structural relations have on genome activity and DNA function through GMOL's unique features and functions, which includes the multi-scale method that can satisfy the users' requirement to not only visualize genome tertiary structure, but also measure it.

# Background

Recent studies have shown that in addition to the genome sequence, the genome's spatial structure also has a tremendous impact on genome activity and DNA function including gene expression and genome stability [1]. Lately, much work has been done in an attempt to decipher such genome structures and several different models have been proposed as probable structures [2, 3, 4, 5, 6, 7]. Moreover, several computational methods have been developed to construct realistic 3D structures of genomes or chromosomes from chromosomal conformation capturing data generated by next generation sequencing techniques [2, 3, 16].

Considering the amount of work in genome structure modeling, a visualization tool that can help researchers visualize and analyze 3D genome structures will undoubtedly benefit genome structure study. Visualization of genome structures is vital to continuous progress in the field because it showcases relationships within the genome that cannot be inferred from sequence information alone. Despite the importance of visualization, not much progress has been made in this area until now. Several tools have been developed for small molecular, such as protein, structure visualization such as Jmol [8], Pymol [9], Chimera [10], etc., but, when it comes to large-scale structures, such as the human genome, there are two limitations that prevent these programs from being effective. First, the Protein Data Bank (PDB) file format is typically used to store molecular structure data for the tools to visualize, however, the standard PDB format was designed for small molecular structures [11] and consequently is not sufficient for storing the vast amounts of data required to visualize genome structures. Second, running tools to load the entire genome structure data so that it is compatible with these programs is a strenuous or impossible task.

To our knowledge only one tool, Genome3D [12], has been specially designed for genome structure visualization. However, Genome3D lacks advanced functions in a few key areas including selection functions and scales amongst others. Here we introduce another genome structure visualization tool named GMOL that adequately improves upon and provides much needed functions, thus successfully filling the need for a genome visualization tool for researchers.

# Implementation

GMOL was developed from Jmol, an open-source Java application that visualizes chemical structures [8]. GMOL adds and modifies several functions to make genome structure visualization possible, sufficient, and GMOL's specific purpose thus differentiating GMOL from Jmol. The added and modified functions, described below, include the scaling system, selection system, sequence querying, measuring system, and new file format.

To visualize and store genome structures, GMOL utilizes a six scale system. The six scales are (listed from lower resolution to higher resolution or from large scale to small scale): genome scale (Gb), chromosome scale (50-100Mb), loci scale (Mb), fiber scale (Kb), nucleosome scale (100b) and nucleotide scale (1b). A smaller scale structure is a component of the next larger scale structure, therefore a larger scale is comprised of the combination of the components of the smaller scale. For example, the genome scale visualizes all the chromosomes; the chromosome scale visualizes all the loci, etc. This multi-scale system is largely inspired by the "fractal globule model" of genome structure as described in [13]. **Figure 1** shows the human genome structure visualized at different scales using data from the modeling structure described in [13]. The different scales can easily be traversed either through the entire current scale, or through a selection of the current scale. For example, the user can view the chromosome scale by viewing one chromosome or the entire collection of all chromosomes.

The backbone of the multi-scale visualization system is toggling between scales. To do this, GMOL utilizes another feature which allows the user to select any unit, at any scale, and scale it up to a lower resolution or down to a higher resolution. By scaling up, the user gets an overview of the location of the selected unit, whereas scaling down gives the user the detailed structure of the selected unit. The visualization at each scale can be dynamically rotated, translated, colored and zoomed in and out.

There are multiple ways in which the user can select a unit or units. One way is called index selection, in which the user can select units by using their index in the current displaying structure. In global scale structure, unit index means chromosome number, while in chromosome/loci/fiber/nucleosome scale unit index means the sequential number of this unit based on genome sequence. Another selection method is via scale information, in which the user selects units by scale information. This method is useful if the user needs to select units within a specified chromo-some/loci/fiber/nucleosome in the current displaying structure when the index in unknown. Lastly, the user can select units using genome sequence information. By specifying a genome sequence location, the corresponding units in current displaying structure will be selected.

In addition to scaling from a selection of units, GMOL can query the selected units into an Ensembl [14] database or a local database to gather genome sequence information about the selection. The integration of JEnsembl [15] with GMOL enables querying of the Ensembl database.

Another feature of GMOL is its measuring capabilities. GMOL allows the user to measure certain selected units in the currently visualized structure. Specifically, GMOL can measure the distance in between any two units in nanometers, and measure the angle formed between any three units in degrees.

With regards to the file format in consideration, currently the standard file format for 3D visualization of biological data is PDB. However, the existing PDB file format is standard for storing protein structures and, therefore, is inadequate for storing genome structure data as genome structure data has a much higher resolution and therefore is much larger. To solve this problem, a new file format, GSS, was designed. Corresponding to our multi-scale system, the GSS format contains the following files (from lower to higher resolution): ".gs.gss" (genome scale), ".cs.gss" (chromosome scale), ".ls.gss" (loci scale), ".fs.gss" (fiber scale), and ".ns.gss" (nucleosome scale). Files of lower resolution store the position of the central point of compartments of the next higher resolution. More specifically, ".gs.gss" files contain the location of the central point of all chromosomes, ".cs.gss" files contain the location of the central point of all loci, ".ls.gss" files contain the location of the central point of all fibers, ".fs.gss" files contain the location of the central point of all nucleosome core particles (NCP), and finally "ns.gss" files contain all the nucleotides in a NCP. Based on this hierarchical organization of the GSS file system, GMOL is able to read and display structures at any resolution according to the user's requirements.

## Results and Discussion

**Functionality of GMOL**

Because GMOL is developed from Jmol, it integrates some existing functions of Jmol. However, the additional functions have been incorporated that prove useful for visualizing genome structures. The functionality and usefulness of the added functions and features are described below. More information on GMOL, including a detailed walkthrough, installation guide, and user guide, can be found in the documentation on the GMOL website.

The multi-scale system of GMOL allows the various resolutions of the structure to be viewed with accuracy and precision. In addition, giving each scale its own file type allows for faster viewing and scaling between scales. The multi-scale system also allows for more total data to be represented by giving each scale its own system of data points. This, in turn, creates a more accurate and reliable genome structure.

The selection system grants flexibility and ease in terms of how units are selected. This makes selection easy and simple as certain selection methods are better suited for certain ranges of data. In addition, users are free to use different selection methods based on their preference.

Flexibility is also represented in GMOL with regards to querying databases. Since GMOL allows querying to a local database or Ensemble, users are free to choose based on their preference and aren't limited.

The measuring system incorporated into GMOL supplies convenient methods to obtain data regarding the genome structure. Via the interface or console, users can measure distances or angles with respect to selected units with ease.

Finally, the unique file type create for GMOL, GSS, allows GMOL to visualize various scales of structures to its fullest ability as the GSS file system grants a higher resolution of visualization and larger amounts of data to cope with the demands of genome structures of which the PDB file system can't provide.

**Visualization Examples of GMOL**

**Figure 2** shows an extracted image of a resulting 3D genome structure visualized in GMOL. The visualization was done in the chromosome scale so each chromosome is visualized in full and with their respective positions to each other. Here, each chromosome within the genome is highlighted with a different color and labeled for identification. The visualization is from a genome modeled from [16] based on Hi-C data [2].

**Figure 3** shows two screenshots taken of GMOL with a visualized genome open. The interface is shown as well as the measurement tool in use. The visualized models are in the genome scale. One of the chromosomes, represented as point in genome scale, is highlighted.

**Applicability of GMOL through Analyzing Genome Structures**

Here, we give an example of how GMOL could be used in a practical situation to analyze the differences between the genome structures of two individuals. As previously mentioned, research has shown that genome spatial structures impacts genome activity and DNA function [1]. This means that the variations in genome structures amongst individuals could account for minor differences such as eye color, but also for more major health concerns such as cancer. By using

GMOL to analyze genome structures, researchers/biologists can quickly spot abnormal sections of the genome and easily scale-up or down to get a more detailed view of the areas of concern.

In **Figure 3**, the genome of Person A is displayed on the left and the genome of Person B is displayed on the right. As shown, the genome structures are almost identical except for the location of Chromosome 1 (highlighted in red for Person A and green for Person B). Assuming Person A is healthy and Person B has been diagnosed with cancer, this difference in positioning of Chromosome 1 should cause concern. By selecting the chromosome unit and scaling down we can get a more detailed view of the structure of Chromosome 1 for each individual (Figure 3).

**Figure 4** highlights the spatial changes in the structure of Chromosome 1 between Person A and Person B. To view these structural differences within the context of the entire genome at the chromosome scale, we simply scale-up to go back to the genome scale and then scale-down to show the global structure at the chromosome level (**Figure 5**).

This short example demonstrates how one might use GMOL in a practical scenario to analyze the differences between two genome structures. For a more comprehensive walkthrough of GMOL's features see the walkthrough on the GMOL website.

**Comparison with Genome3D**

GMOL improves upon Genome3D [14] by supplying some important and needed features that are necessary for adequate genome visualization, of which Genome3D lacks. One way this is demonstrated is through the available scales offered, in which Genome3D displays genome structures at three scales: Giant Loop, Fiber and Nucleosome. GMOL's multi-scale system utilizes six scales: genome, chromosome, loci, fiber, nucleosome, and nucleotide. These additional scales allow a more detailed view of genome structures that cannot be achieved with Genome3D. GMOL also implements multiple selection functions (based on index, based on scale information, based on sequence) whereas Genome3D only allows selection based genome location. Having multiple selection functions enables intuitive selection of any portion of the genome structure from any scale. Furthermore, GMOL supports distance and angle measurement functions that Genome3D lacks. With regards to sequence querying, both Genome3D and GMOL support it, but GMOL supports querying from Ensembl and from a local database whereas Genome3D supports querying only from a local database. Finally, GMOL allows the user to write custom scripts and commands so that they may extend the functionality of GMOL to needs specific to their project. Genome3D does not implement this feature.

Ultimately, GMOL and Genome3D perform the same basic functions, however GMOL allows for more detailed analysis of genome spatial structures in several key areas which allows researchers to achieve better answers regarding structural relationships. The main differences between GMOL and Genome3D are summarized in **Table 1**.

**Future Development of New Features**

A number of future developments of new features of GMOL are planned for implementation and integration in the near future. One such feature is the integration of additional databases of which to export sequences from GMOL to. Currently, GMOL supports querying sequences to Ensemble and a local database. Additional databases being integrated include UCSC Genome browser [17], ENCODE [18], and Uniprot [19]. Another planned feature to include is a function that allows the selection of two points, of which to then be visualized for comparison. This function will allow an easier and better method to compare two sections of the genome. A third feature planned for future development is the integration of a function to view the sequence of a selected point. Such function would be convenient with regards to getting the sequence of only a selection.

# Conclusion

The recent development and research in genome nature indicates the significance of 3D genome structures as well as genome sequence. The genome's structure provides a key contribution to certain genome activity and DNA functions including gene expression and genome stability [1]. The implications of such findings suggest that much work is needed to figure out 3D structures of genomes. One vital step in the process of studying 3D genome structures is the visualization process which turns the coordinates of a generated structure into a 3D, interactive model.

GMOL is an application designed to sufficiently perform this step of the research process by effectively visualizing genome tertiary structures. GMOL does this through its immense array of features and functions including its multi-scale system that allows visualization of six chronological resolutions. GMOL also successfully fulfills its goals through its multi-selection system, measurement capabilities, options of sequence querying, and new file format system.

Therefore, through GMOL, researchers can better analyze their genomics data. By using GMOL to visualize the genome, researchers may see patterns or other structural relationships that are not evident in their data alone. By utilizing six different scales, GMOL allows for a level of detail that cannot be obtained by any other currently released program. That combined with GMOL's other unique capabilities set it apart in the marketplace for genome visualization software.

## Availability and requirements

**Project name:** GMOL

**Project homepage:** http://sourceforge.net/projects/gmol/

**Operating Systems:** Windows, Mac, Linux

**Programming language:** Java

**License:** GNU Library or Lesser General Public License version 3.0 (LGPLv3)

**Other requirements:** Java 1.7 or higher

# Figures

## Figure 1 - Visualization of Human Genome Structure of Different Scales

The structure at each level is visualized in a dynamic fashion such that it can be rotated, translated, colored, and zoomed in and out.

**Figure 2 – Image of Visualized Genome on Chromosome scale using GMOL.**

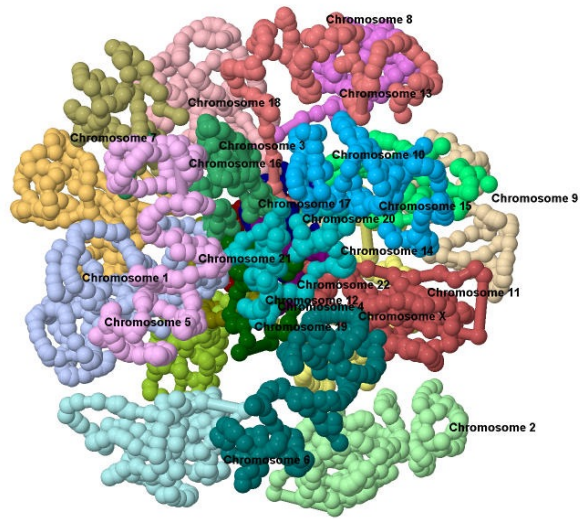

Extracted image of a resulting 3D genome structure visualized in GMOL, in the chromosome scale. Here each chromosome within the genome is highlighted with a different color and labeled for identification. The visualization is from a genome modeled from [16].

**Figure 3 – Comparison of Genome Structures at Genome Scale**

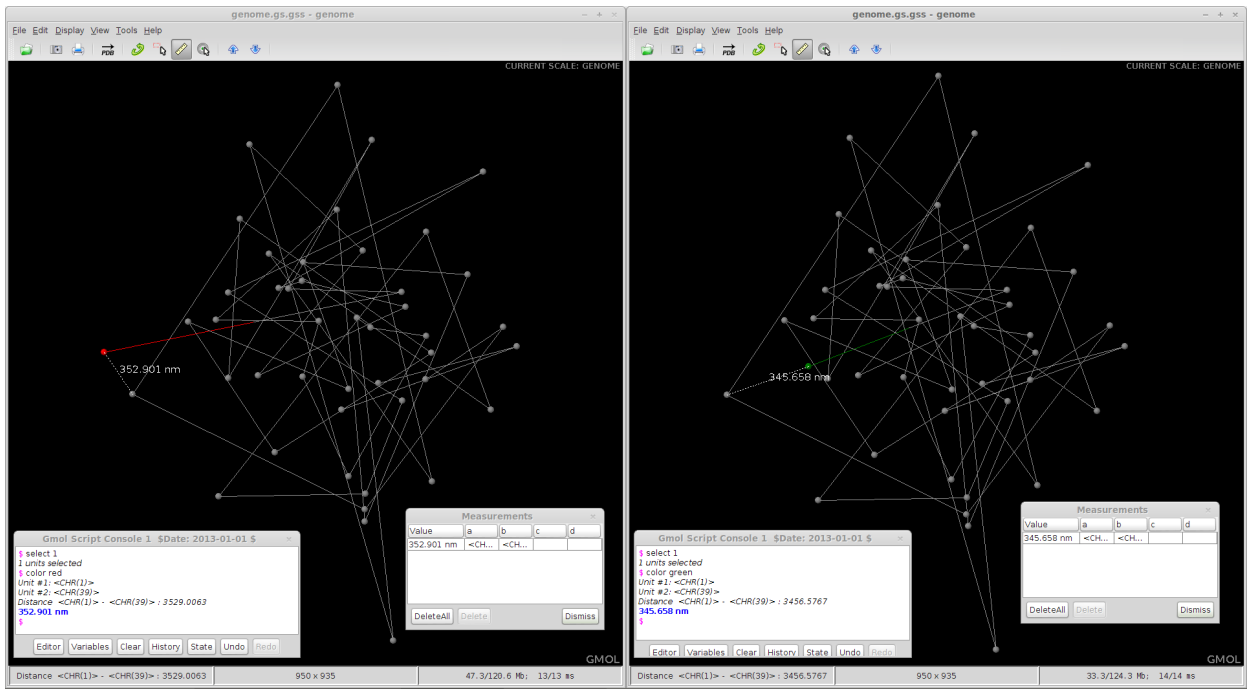

Side by side screenshots of GMOL visualizing the genome of two models at genome scale with Person A on the left and Person B on the right. The difference in position of Chromosome 1 between the two models is highlighted.

**Figure 4 – Comparison of Chromosome Structures at Chromosome Scale**

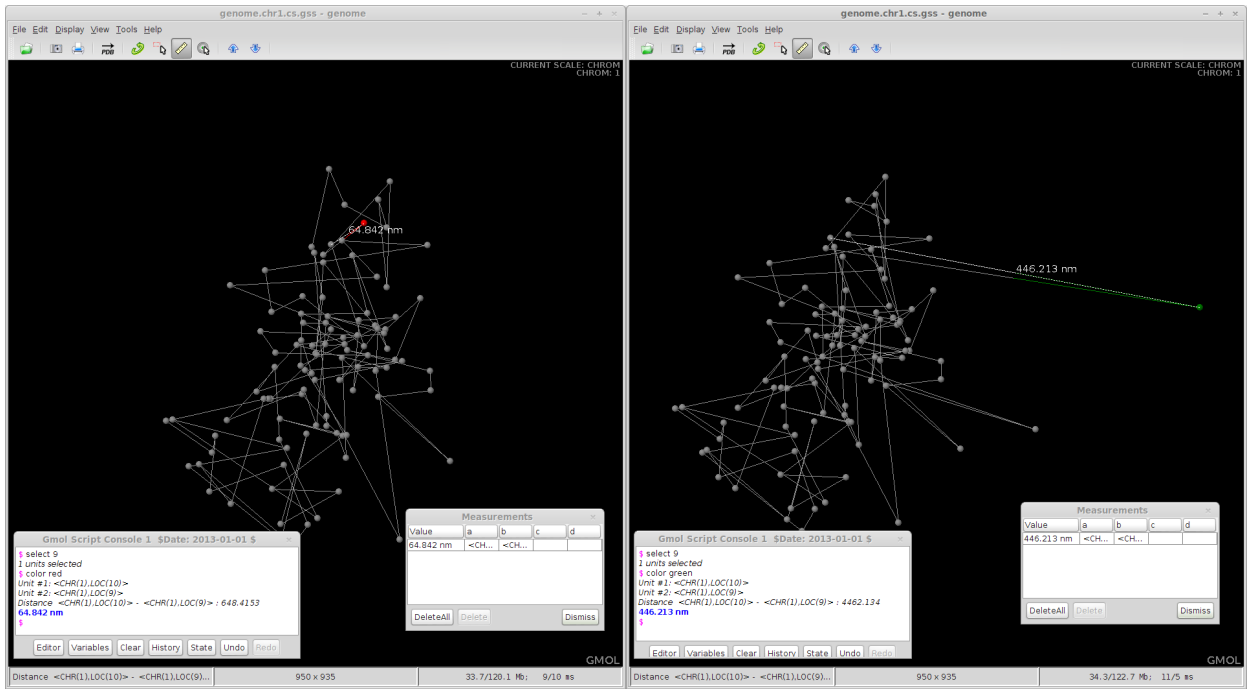

Side by side screenshots of GMOL visualizing chromosome 1 of two models at chromosome scale with Person A on the left and Person B on the right. The structural differences within Chromosome 1 are highlighted.

**Figure 5 – Comparison of Chromosome Structures at Genome Scale**

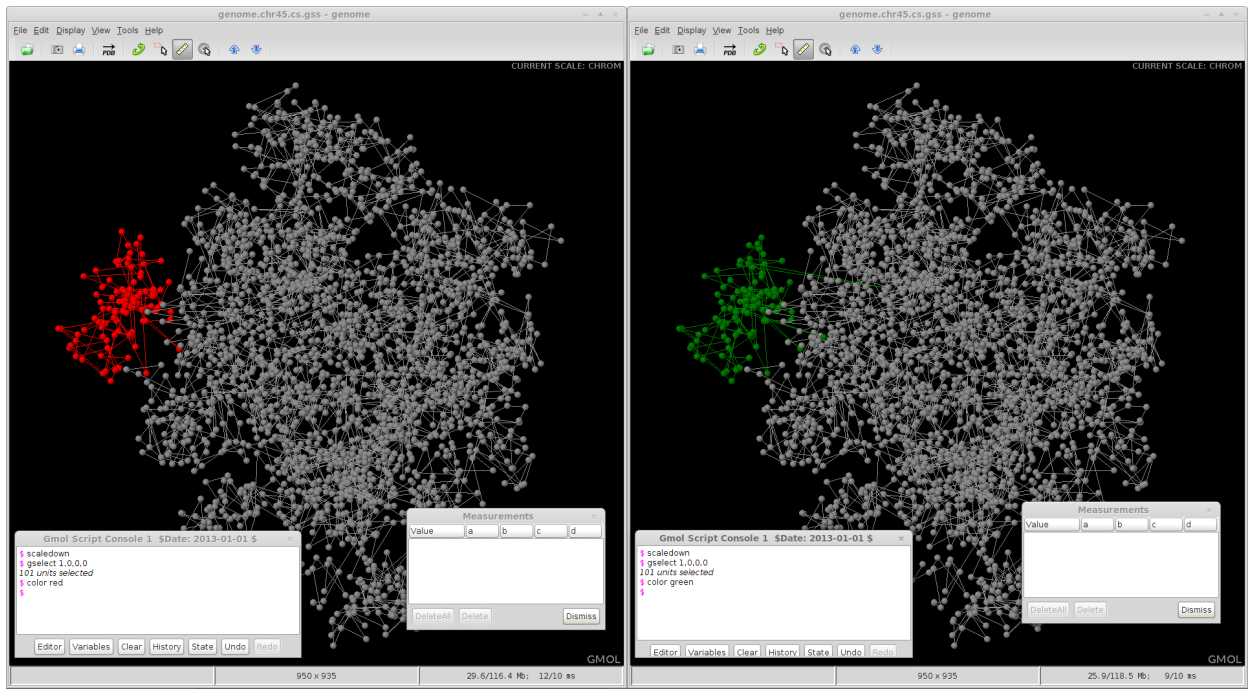

Side by side screenshots of GMOL visualizing chromosomes of two models at the genome scale with Person A on the left and Person B on the right. The structural differences of Chromosome 1 within the context of the entire genome are highlighted.

# Tables

### Table 1 - Comparison of GMOL and Genome3D

| Functions | GMOL | Genome3D |
|---|---|---|
| *Select Function* | 1. select based on index<br>2. select based on scale information<br>3. select based on sequence | Select only based on genome location |
| *Measurement* | Supported | Not Supported |
| *Sequence querying* | 1. query from Ensembl<br>2. query from local database | Only query from local database |
| *Scripts/Commands* | Supported | Not Supported |
| *Visualization Scales* | Genome<br>Chromosome<br>Loci<br>Fiber<br>Nucleosome<br>Nucleotide | Giant Loop<br>Fiber<br>Nucleosome |